\documentclass[aps,prd,twocolumn,nofootinbib,floatfix]{revtex4-1}
\usepackage{setspace}
\usepackage{amsthm}
\theoremstyle{definition}

\usepackage{graphicx}
\usepackage{dcolumn}
\usepackage{multirow}
\usepackage{subfigure}
\usepackage{times,mathptm}
\usepackage{braket}
\usepackage{float}
\usepackage{color}
\usepackage{amsmath,amsfonts}
\usepackage{mathptmx}
\usepackage{mathrsfs}
\usepackage{bbm}
\usepackage{bm}
\usepackage{xfrac}

\newcommand{\beq}{\begin{equation}}
\newcommand{\eeq}{\end{equation}} 
\newcommand{\bea}{\begin{eqnarray}}
\newcommand{\eea}{\end{eqnarray}}

\newcommand{\Sc}{S$_\text{c}$}

\renewcommand{\d}{\delta}

\newcommand{\p}{\phi}
\renewcommand{\b}{\beta}

\newcommand{\tr}{\text{Tr}}

\newcommand{\vx}{{\vec{x}}}
\newcommand{\vy}{{\vec{y}}}

\newcommand{\m}{\mu}
\newcommand{\q}{\overline{q}}
\newcommand{\g}{\gamma}

\newcommand{\dg}{\dagger}
\newcommand{\non}{\nonumber}

\newcommand{\rf}[1]{(\ref{#1})}
\newcommand{\ra}{\rightarrow}

\renewcommand{\vec}[1]{\bm #1}
\usepackage{ulem}

\begin{document}

\title{The Higgs phase as a spin glass phase in D=5 dimensional SU(2) gauge Higgs theory} 

\bigskip
\bigskip

\author{David Ward}
%\singlespacing
\affiliation{Physics and Astronomy Department \\ San Francisco State
University  \\ San Francisco, CA~94132, USA}
\bigskip
\date{\today}
\vspace{60pt}
\begin{abstract}

\singlespacing
 
    According to recent work of Greensite and Matsuyama, the Higgs phase of a gauge Higgs theory is distinguished from the
confinement and massless phases by the spontaneous breaking of a global center subgroup of the gauge group, and by confinement type.  This is contrary to the notion that there is no essential distinction between the Higgs and confinement phases when the Higgs field is in the fundamental representation of the gauge group.  Although this new symmetry breaking order parameter has been investigated in $D=4$ dimensions, there is so far no check in a non-abelian gauge theory containing a massless as well as confinement/Higgs phases, where the prediction is that the symmetry breaking order parameter will show transition lines separating the massless to Higgs and confinement to Higgs phases, but not the massless to confinement phase.  In this work we map out the phase structure of the $D=5$ dimensional model, according to both the symmetry breaking parameter and thermodynamic observables, and check the assertion regarding the massless to confinement phase.

\end{abstract}

\maketitle

\singlespacing
%\begin{widetext}
%\section{\label{intro}Introduction}++

\section{\label{sec:Intro} Introduction}

    One often hears that there is no true distinction between the Higgs and confinement phases of a gauge Higgs theory, when the Higgs field is in the fundamental representation of the gauge group.  There are a number of strong arguments which support that assertion.  First, there is
no obvious gauge invariant order parameter, such as a Polyakov line, which would distinguish the two phases.  Secondly, it is known
from Elitzur's theorem that a local symmetry cannot break spontaneously, and it therefore seems erroneous to describe the Higgs phase
as a phase of spontaneously broken gauge symmetry.  Third, there is the work of Osterwalder and Seiler \cite{Osterwalder:1977pc}, Banks and Rabinovici \cite{Banks:1979fi}, and
Fradkin and Shenker \cite{Fradkin:1978dv}, which tells us that the Higgs and confinement phases cannot be entirely isolated from one another in the phase diagram by a thermodynamic transition.  Finally, the asymptotic particle states of, e.g., an SU(2) gauge Higgs theory are created by local color singlet operators, in both the Higgs and confinement regions of the phase diagram \cite{Frohlich,tHooft:1979yoe,Banks:1979fi}.  Of course it is always possible to fix to some gauge which leaves a global subgroup of the gauge group unbroken, and then there may be a transition between a region where this global symmetry is unbroken, and a region where it is spontaneously broken.  But it was shown in \cite{Caudy:2007sf} that such transition lines depend on the gauge choice, which makes a distinction of this kind ambiguous at best.   
	
     Recently Greensite and Matsuyama \cite{Greensite:2020nhg} have argued that there is, in fact, an essential distinction between the Higgs and confinement 
phases, and that these two phases differ in two ways:
\begin{itemize}
\item {\bf Confinement type}.  When the asymptotic spectrum contains only color neutral particles, this is called color (or ``C'') confinement,
and it is found in both the Higgs and confinement phases.  However, in the confinement phase there exists a stronger version of
confinement, described by these authors as separation of charge (or ``\Sc'') confinement, which is associated with the formation
of metastable flux tubes, and linear Regge trajectories.
\item {\bf Symmetry}.  The Higgs phase is characterized as a kind of spin glass, in which a global center subgroup of the full gauge group
is spontaneously broken.  The order parameter for this symmetry breaking is closely analogous to the one introduced by Edwards and Anderson \cite{Edwards_Anderson} in the spin glass context, and does not make any reference, even implicitly, to a gauge choice.
\end{itemize}
	
     In the next section we will explain these claims in a little more detail, leaving their justifications to the cited reference.  The purpose
of the present work is to compute the symmetry breaking (``Edwards-Anderson'') order parameter in SU(2) gauge Higgs theory in $D=5$ dimensions, along with more conventional thermodynamic observables. There are two motivations
for going to five dimensions, both of which are related to the existence of a massless phase in the five dimensional lattice gauge Higgs theory, a phase which is absent in four dimensions.  The first motivation is to check the prediction that both the confinement and massless phases are symmetric with respect to the global center subgroup of the gauge group, and to see whether the order parameter sensitive to this symmetry will remain zero across the transition between these two phases. The second is to investigate whether and where the Edwards-Anderson order parameter coincides with thermodynamic transitions to the Higgs phase, particularly from the massless to the Higgs phase.  That comparison has been carried out in the $D=4$ theory for the Higgs to confinement transition, but not for the massless to Higgs transition.  It is of course understood that SU(2) gauge Higgs theory is non-renormalizable in $D=5$ dimensions, but the continuum limit of this theory, taking the lattice spacing to zero, is of no concern here; we are simply concerned with the phase diagram of this theory in the space of coupling constants at some arbitrary but finite lattice spacing.  

In connection with the Higgs to confinement transition, it should be noted that we already know of examples where there exist physically distinct phases of many-body systems which are not separated by a thermodynamic transition.  One example is the roughening transition in Yang-Mills theory. In the rough phase, the width of flux tubes grows logarithmically with quark-antiquark separation, and the static quark potential contains a $1/r$ term of stringy origin. Neither of these features hold outside the rough phase. Another example is the Kertesz line \cite{Kertesz} in the Ising model in an external field , which has to do with a percolation transition in the random cluster
formulation of the model.  And a third example is the chiral symmetry resoration transition for adjoint fermions \cite{Karsch:1998qj}.
	
     Section 2, below, contains a brief exposition of the symmetry breaking order parameter, and the closely related property of separation-of-charge confinement. Section 3 is a discussion of our procedure for locating symmetry breaking and thermal transitions, ending with a
phase diagram for the $D=5$ dimensional theory which incorporates both types of transitions.  Conclusions are found in the final section 4.

\section{Symmetry and \Sc \ confinement}

     In a gauge Higgs theory, with the scalar field in the fundamental representation, there always exists a symmetry which transforms
 the scalar, but not the gauge field.  This is the global center subgroup of the gauge group, with transformations $g(x) = z \in Z_N$.
 The SU(2) group is special, in that the global symmetry group is much larger than $Z_2$.  In SU(2) lattice gauge theory, imposing for simplicity the unimodular constraint $|\phi|=1$, the action can be written in the form
\bea
 S &=&  S_W[U]+S_H[U,\phi] \non \\
    &=& -\beta\sum \frac{1}{2}\tr[U_{\mu}(x)U_{\nu}(x+\hat{\mu})U_{\mu}^\dg(x+\hat{\nu})U^\dg_{\nu}(x)] \non \\
    & & -\gamma \sum \frac{1}{2}\tr[\phi^{\dg}(x)U_{\mu}(x)\phi(x+\hat{\mu})] \ ,
\eea    
 with $\p(x)$ taking values in the SU(2) subgroup.    This theory is invariant under the global transformations
 \beq
        \phi(x) \ra \phi(x) R
 \eeq
 where $R \in SU(2)$ is sometimes referred to as ``custodial symmetry,'' and is a term we also use here.  Note that $R$ contains $Z_2$
 as a subgroup, and this subgroup cannot be distinguished from transformations belonging to the global center subgroup of the gauge group, which also does not affect the gauge field.  The claim is that the Higgs phase corresponds to spontaneous breaking of this subgroup.
 \footnote{In the literature of the electroweak theory, the ``breaking'' of the gauge symmetry refers to a gauge choice which leaves a global
SU(2) symmetry unfixed, and then the remaining global symmetry for SU(2) is SU(2)$_\text{global} \times R$.  In that case
$\p$ may develop a vacuum excitation value which breaks SU(2)$_\text{global} \times R$ down to a diagonal subgroup.  That is not
exactly what we are talking about here since,  as was already mentioned, the line of spontaneous symmetry breaking in this approach is gauge dependent.}

   Since no gauge is fixed, $\langle \phi \rangle$ must vanish, and the question is what order parameter (necessarily gauge invariant) can detect the spontaneous symmetry breaking of the custodial symmetry.  According to \cite{Greensite:2020nhg}, the appropriate order parameter $\Phi$, which has strong similarities to the Edwards-Anderson order parameter \cite{Edwards_Anderson} for a spin glass, is defined in the following way.  The order
parameter $\Phi[U]$ is a gauge-invariant functional of the gauge field $U_i(\vx,t=0)$ on the time slice $t=0$.  Then
\bea   
       \langle \Phi[U(0)] \rangle &=& {1\over Z} \int DU_\m D\phi ~ \Phi[U(0)] e^{-S_E} \non \\
       \Phi[U(0)] &=& {1\over V_3} \sum_{\vx} | \overline{\phi}(\vx;U(0))| \non \\
       \overline{\p}(\vx;U(0))  &=&    {1\over Z_s} \int  D\phi DU_0 [DU_i]_{t\ne 0} ~~ \phi(\vx,t=0) e^{-S_E} \non \\
       Z_s &=&  \int  D\phi DU_0 [DU_i]_{t\ne 0} ~ e^{-S_E} \ ,
\label{Phi}
\eea
and $V_3$ is the 3-volume of a time slice.  The claim is that in the $V_3 \ra \infty$ limit, $\langle \Phi \rangle = 0$ in the confinement
and massless phases, while $\langle \Phi \rangle \ne 0$ in the Higgs phase.  The massless phase in $D=5$ dimensions is isolated from the confinement and Higgs phases in the phase diagram by a line of thermodynamic transitions, so we can determine its location accurately, by standard methods.  Part of the motivation of this paper is to check the assertion that $\langle \Phi \rangle = 0$ throughout the massless phase.

    As in any discussion of spontaneous symmetry breaking, one must formally add a symmetry breaking term which is taken to zero after the
taking the infinite volume limit.  In numerical simulations this nicety is of little practical importance, and will not be needed here.  For a full 
discussion of the symmetry breaking term, as well as the relation of $\Phi$ to the Edwards-Anderson order parameter and to custodial symmetry,  we refer the reader to ref.\ \cite{Greensite:2020nhg}.   One question, however, which cannot be left to a reference, is the physical distinction between the Higgs and confinement phases in SU(2) gauge Higgs theory.  This is the difference between color (C) confinement and separation-of-charge (\Sc) confinement, already alluded to.  

    There is no need to dwell on C confinement, this simply means that all asymptotic particle states are color neutral.  A color charged
particle is the source  of a long-range color electric field, and this is ruled out if the theory is massive.  In the absence of a massless sector,
both the Higgs and confinement phases are C confining.  What distinguishes the Higgs and confinement phases physically is the tendency,
in the confinement phase, to form metastable flux tubes.  It is these flux tubes that are responsible for the linear Regge trajectories
of QCD.  Of course long flux tubes will decay by string breaking.  The question is whether, if such breaking effects were somehow excluded,
the energy of a quark-antiquark pair would diverge with quark separation.  There is of course no way to exclude string breaking experimentally, but it is no problem to do this theoretically.  The idea is consider static quark-antiquark states of the following form:
\beq
          |\Psi_R\rangle  =  \q^a(\vx) V^{ab}(\vx,\vy;U) q^b(\vy)  |\Psi_0\rangle \ ,
\eeq
where $R=|\vx-\vy|$,  $a,b$ are color indices, $V(\vx,\vy;U)$ transforms bi-covariantly at points $\vx,\vy$, and $\Psi_0$  is the vacuum state.  We consider the energy expectation value above the vacuum energy ${\cal E}_0$
\beq
             E_V(R)  =  {  \langle \Psi_R | H |  \Psi_R\rangle \over \langle \Psi_R | \Psi_R\rangle } - {\cal E}_0 \ .
\eeq
A theory is \Sc\ confining iff $E_V(R) \ra \infty$ for any bicovariant functional $V(\vx,\vy;U)$ whatsoever, as $R\ra \infty$.  The crucial point is that the functional $V$ should depend {\it only} on
the gauge field, and not on any matter fields that may be in the theory.  This is the ``theory'' approach to excluding string broken states.
Of course when $\Psi_R$ evolves in Euclidean time, the string can break.  The question is whether there exists some choice of $V$, independent of the matter field, such that the $E_V(R)$ goes to a finite value at infinite time separation.  If so, the theory is not \Sc\
confining.  Speaking a little loosely, the question is whether the system ``wants'' to form a flux tube, and this desire is thwarted by
string breaking, or whether the system can avoid infinite energy at infinite separation without resorting to a string-breaking mechanism; i.e.\ without using the matter field to neutralize the charge of the static source .
A system which forms metastable flux tube states is, of course, a system characterized by linear Regge trajectories, and this is the
physical distinction between an \Sc\ confining phase, and a Higgs or massless phase.  According to ref.\ \cite{Greensite:2020nhg}, a system in which custodial symmetry is spontaneously broken is in the C confinement phase, i.e.\ the Higgs phase.  If the symmetry is not broken then the theory is in either a confinement or a massless phase.

    It is that last statement that we would like to check in this article.  SU(2) gauge Higgs theory does not have a massless phase in $D=4$
dimensions, but there is such a phase in $D=5$ dimensions.  For the confinement to Higgs transition we can only check whether the
transition in $\Phi$ coincides with a thermodynamic transition, where such a transition exists; apart from that the boundary between the
confined and Higgs phases is located by $\Phi$ alone.  The massless phase, however, is isolated by thermodynamic transitions from both
the confined and Higgs phases.  So in the $D=5$ theory we are in a position to check that (i)
$\langle\Phi\rangle=0$ everywhere in the massless phase; (ii) there will be a transition to  $\langle\Phi\rangle>0$ across the massless-to-Higgs thermodynamic transition line; (iii)  there will be no transition in the order parameter across the massless-to-confinement thermodynamic  transition line.

\section{Procedure and Results}

    Most of our numerical simulations were carried out on an $8^5$ lattice volume.  The calculation of the $\Phi$ order parameter is
quite lengthy, involving a ``Monte Carlo within a Monte Carlo'' procedure, to be described below, which limits us to a fairly small
lattice extension in five dimensions.  Of course, larger volumes might be accessible with a massively parallel computation, but that was not implemented in this work, since good results are obtained even with the small lattice we have used here.

\subsection{Thermodynamic transition points}

    We begin with the location of thermodynamic transitions.  There are three transition lines of interest:
\begin{itemize}
\item The line separating the massless and confining regions.
\item The line separating the massless and Higgs regions.
\item A line with an endpoint, lying between the confinement and Higgs regions.  There must be an endpoint to this line, as we
know from  \cite{Osterwalder:1977pc,Banks:1979fi,Fradkin:1978dv}.
\end{itemize}

\begin{figure}[t!]
\includegraphics[scale=.5]{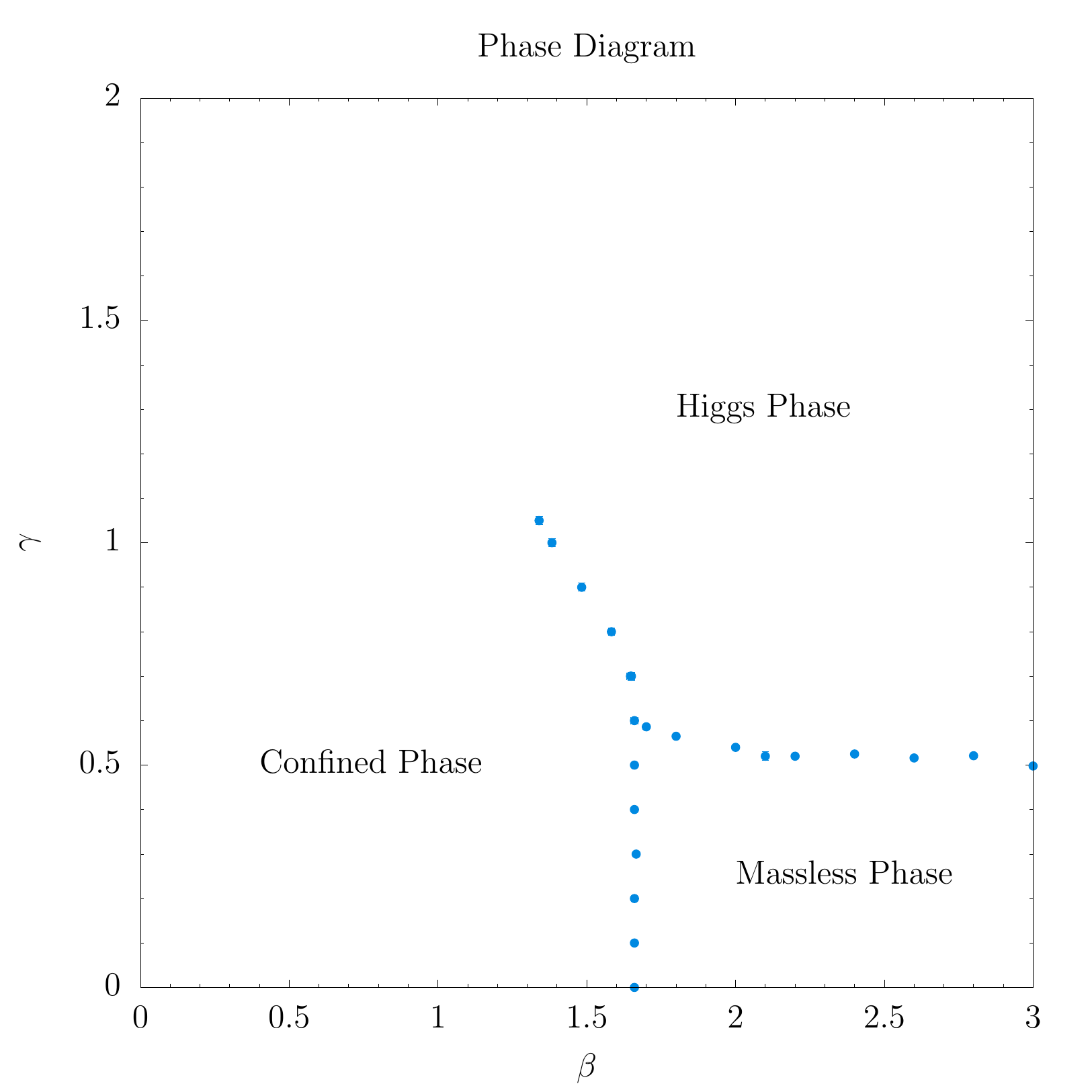}
\caption{Phase diagram of SU(2) gauge Higgs theory in five dimensions, showing only thermodynamic transitions.}
\label{thermo}
\end{figure}

\begin{figure*}[htbp]
\subfigure[~]{
   \includegraphics[scale=0.5]{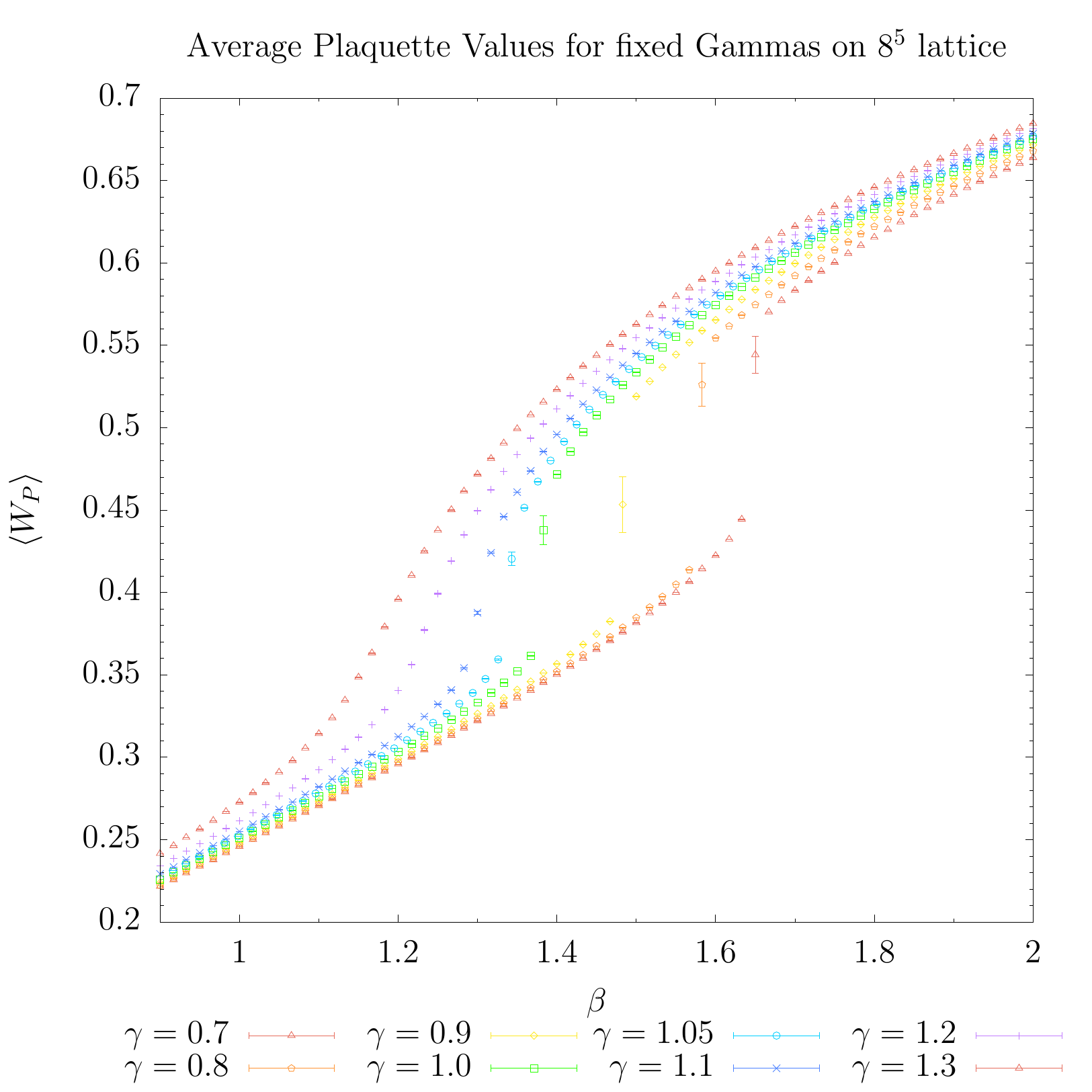}
   \label{fig:gambetpl}
}
\subfigure[~]{
   \includegraphics[scale=0.5]{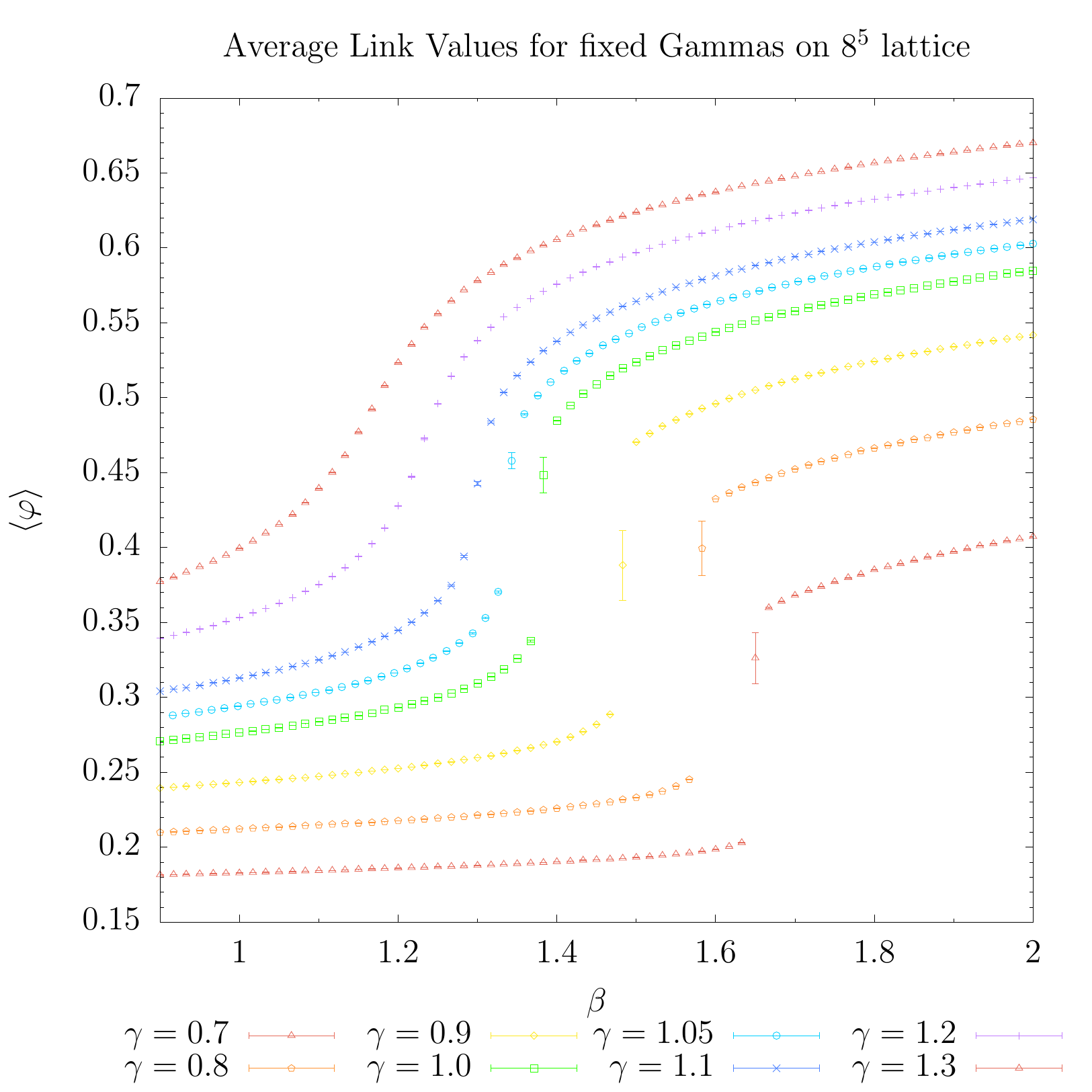}
   \label{fig:gambetsc}
}
\caption{A line of thermodynamic transition from the confinement to the Higgs phase is determined from jumps in both (a) the plaquette
and (b) the gauge invariant link $\langle\tr[\phi^\dg(x)U_\m(x)\phi(x+\hat{\m}])\rangle$ vs.\ $\b$ at fixed $\g$.  These transitions are
first order, and the transition line has an endpoint, as expected from the Osterwalder-Seiler theorem \cite{Osterwalder:1977pc}.  A rough
estimate of the location of that endpoint, just from these figures, is around $\b \approx 1.3, \g \approx 1.1$.}
\label{thermo_conftoHiggs}
\end{figure*}

\begin{figure}[htbp]
   \includegraphics[scale=0.5]{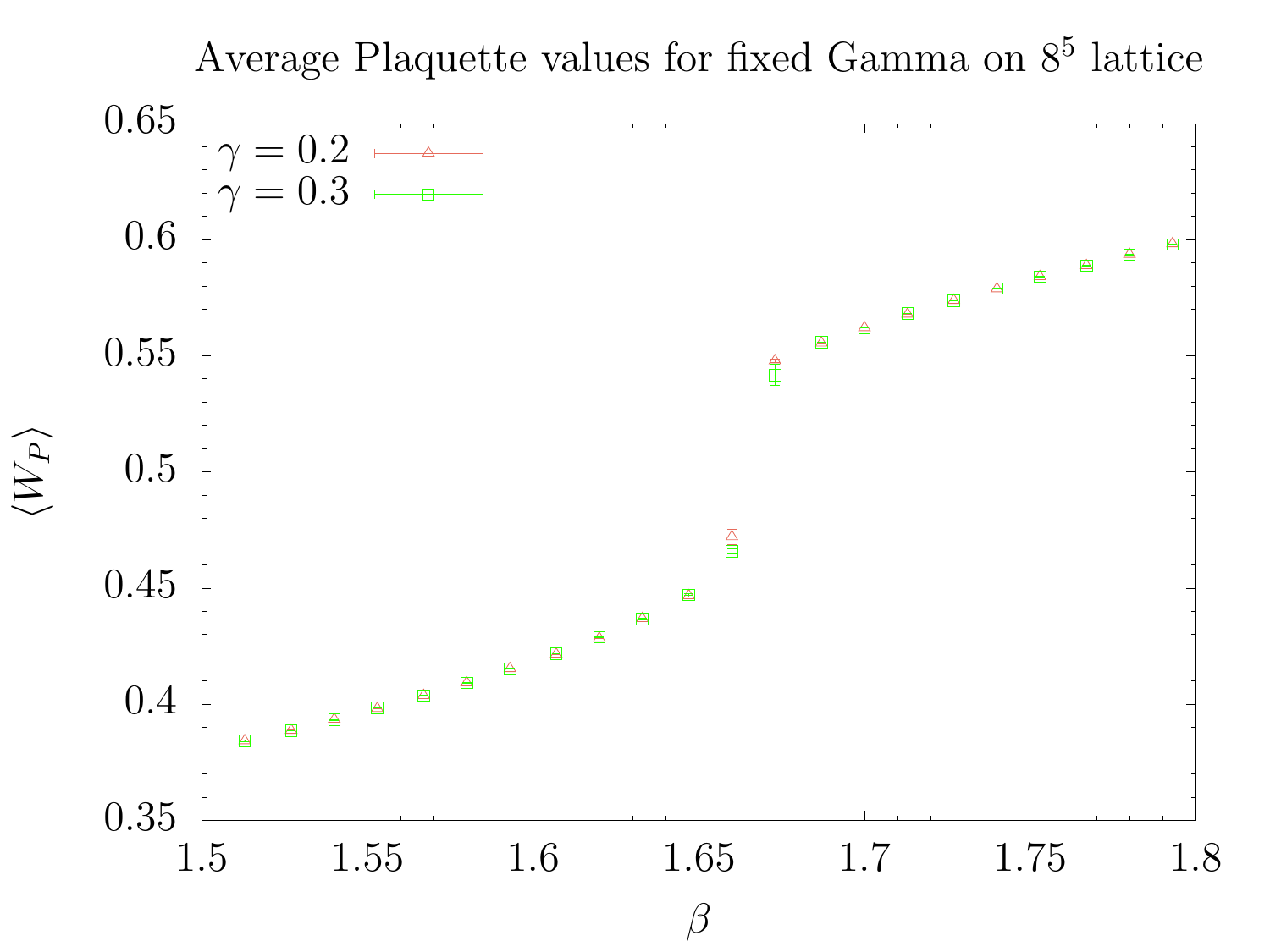}
\caption{The confinement to massless phase transition is signaled by a jump in the plaquette expectation value vs.\ $\b$ at fixed $\g$.  This appears to be a first order transition.}
   \label{thermo_m0toconf}
\end{figure}
 
 \begin{figure}[htbp]
   \includegraphics[scale=0.5]{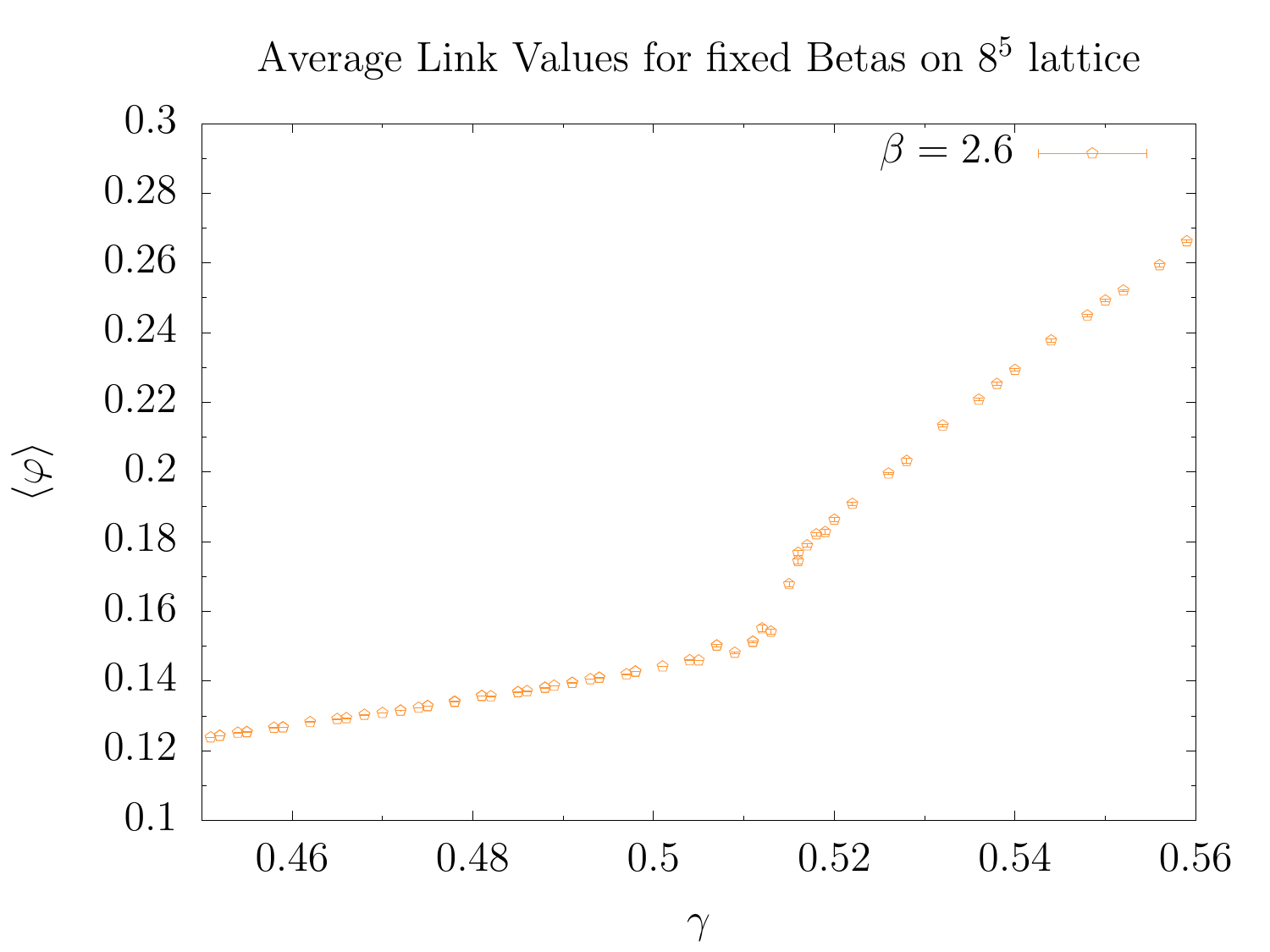}
 \caption{The massless to Higgs phase transition is determined by a ``kink'' in a plot of the
gauge invariant link $\langle\tr[\phi^\dg(x)U_\m(x)\phi(x+\hat{\m})]\rangle$ vs.\ $\g$ at fixed $\b$, and is presumably a continuous transition of some kind.  Similar behavior in the massless to Higgs transition has been seen in the abelian Higgs model \cite{Matsuyama:2019lei}.}   \label{thermo_m0toHiggs}
\end{figure}
	
We determine these thermodynamic points by identifying apparent non-analytic behavior in the plaquette variable and the
gauge-invariant link variable
\beq
  \varphi = \langle\tr[\phi^\dg(x)U_\m(x)\phi(x+\hat{\m})]\rangle \ ,
\eeq
The resulting phase diagram is shown in Fig.\ \ref{thermo}; it is very similar to that of the five dimensional SU(3) gauge Higgs theory reported by Beard et al.\ \cite{Beard:1997ic}.\footnote{ Pure SU(2) gauge theory in five dimensions, with one small dimension compactified toroidally or via orbifolding, has been investigated intensively by Knechtli and co-workers, see e.g.\ \cite{Hollwieser:2018nrh,Knechtli:2016pph} and references therein.
The motivation is that the compactified pure gauge theory in $D=5$ dimensions would generate an effective gauge Higgs theory in $D=4$ dimensions.  Of course in the present article we have an uncompactified theory with a Higgs field introduced from the start, and the motivations are rather different. It might still be of interest to apply our approach to the effective 4D theory generated by the compactified theory.}  A sample of data for the link and plaquette expectation values, used to determine transition points between the confinement and Higgs phases  is shown in Fig.\ \ref{thermo_conftoHiggs}.  Here the link and plaquette values are plotted vs.\ $\beta$ at various $\gamma$, and the transition points are identified by an apparent discontinuity in those observables.  Of course, the designation ``confinement to Higgs'' is at this stage tentative, pending the delineation of these phases by the $\Phi$ order parameter.  There are no such ambiguities in designation for the massless to confinement transition, which is a strong first order transition also signaled by an apparent discontinuity in the plaquette variable, see Fig.\ \ref{thermo_m0toconf}.  The massless to Higgs transition 
is a much weaker transition, signalled by a ``kink'' in the plot of $\varphi$ vs.\ $\g$ at fixed $\b$, e.g.\ in Fig.\ \ref{thermo_m0toHiggs}.  A kink of precisely this sort, locating the massless to Higgs transition and determined with
substantially larger lattices and better statistics, is found also in a four dimensional abelian Higgs theory \cite{Matsuyama:2019lei}.

\subsection{Symmetry breaking transition points}

\begin{figure*}[h]
\subfigure[~]{
\includegraphics[scale=0.5]{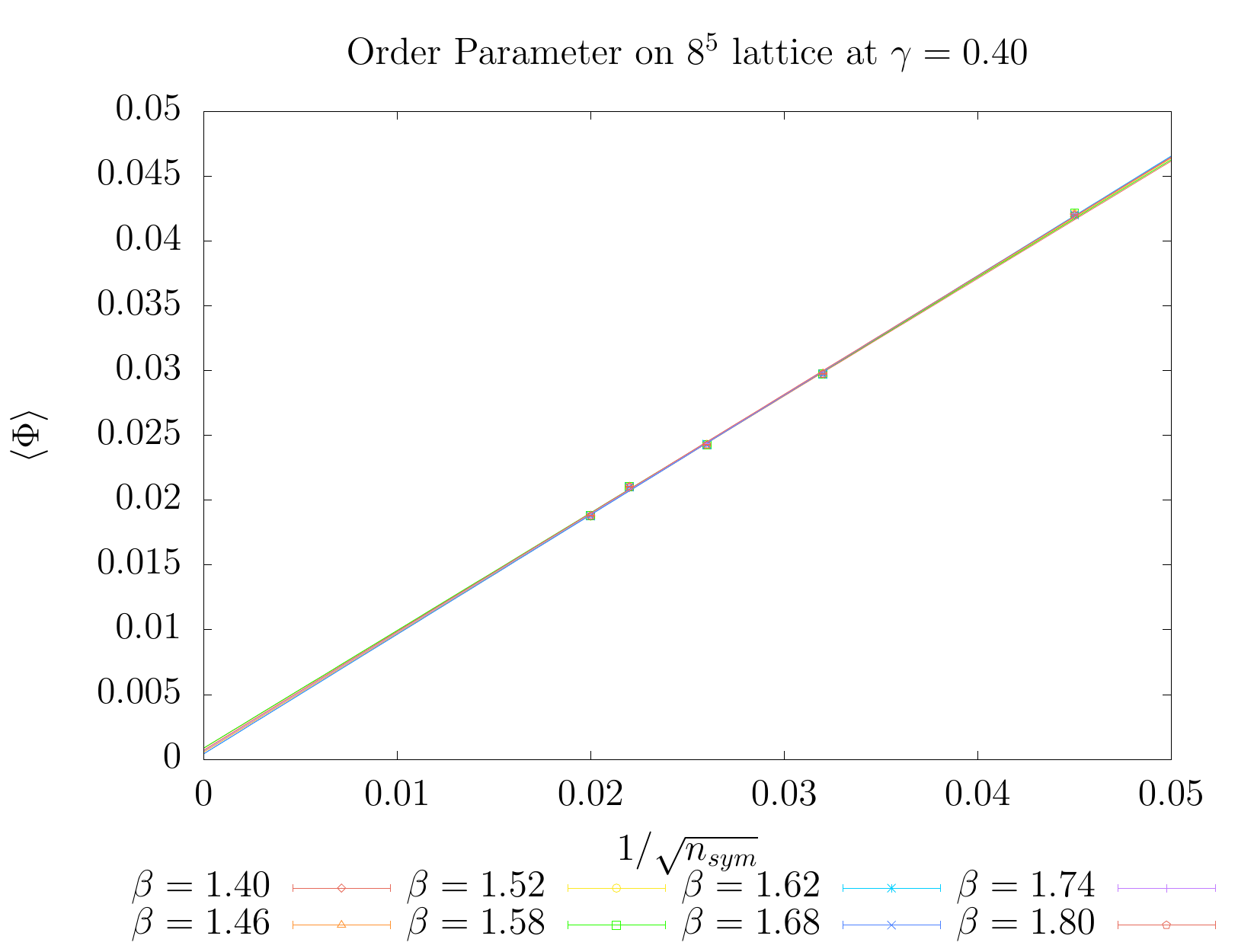}
\label{Phim0toconf}
}
\subfigure[~]{
 \includegraphics[scale=0.5]{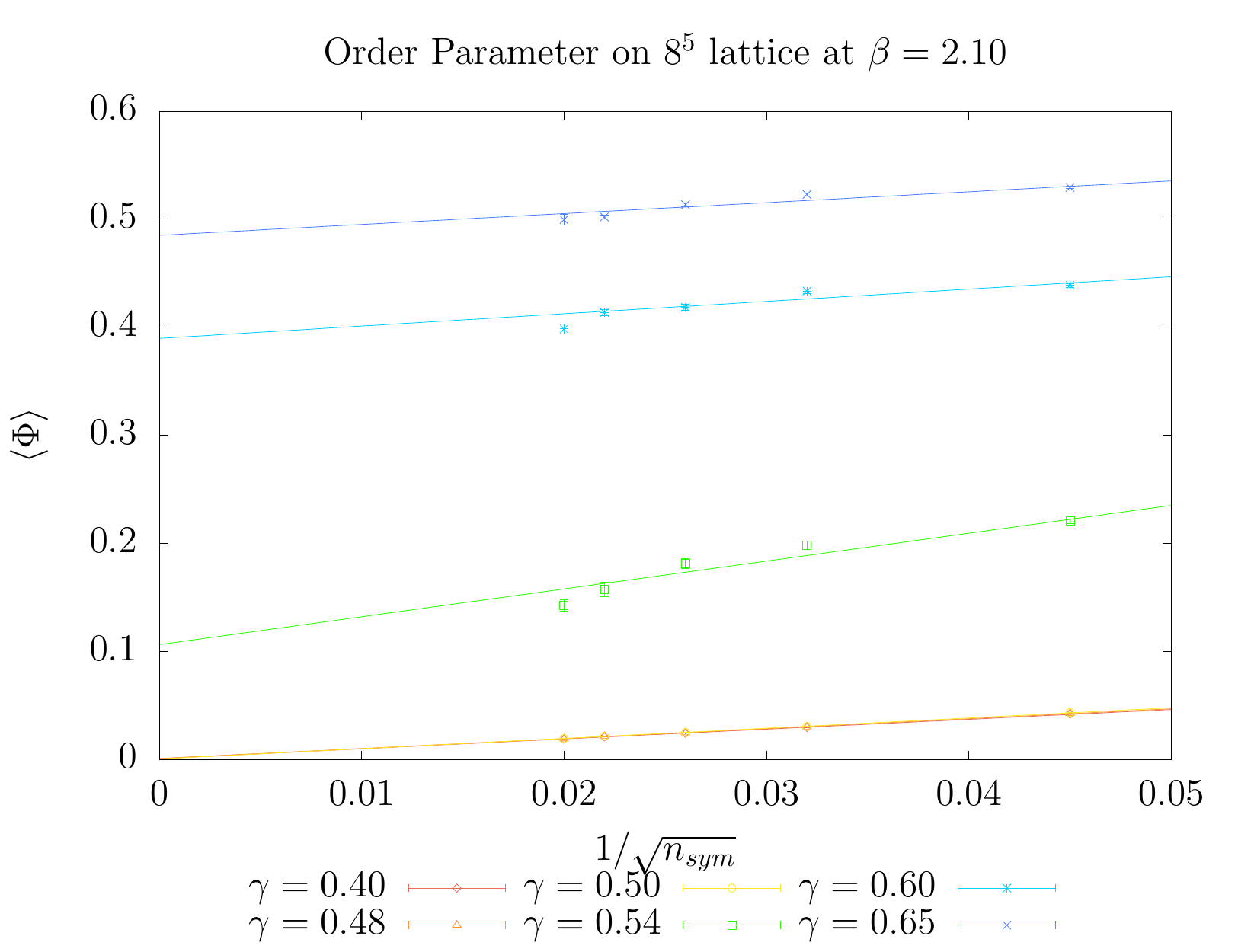}
 \label{Phim0toHiggs}
 }
\subfigure[~]{ 
 \includegraphics[scale=0.5]{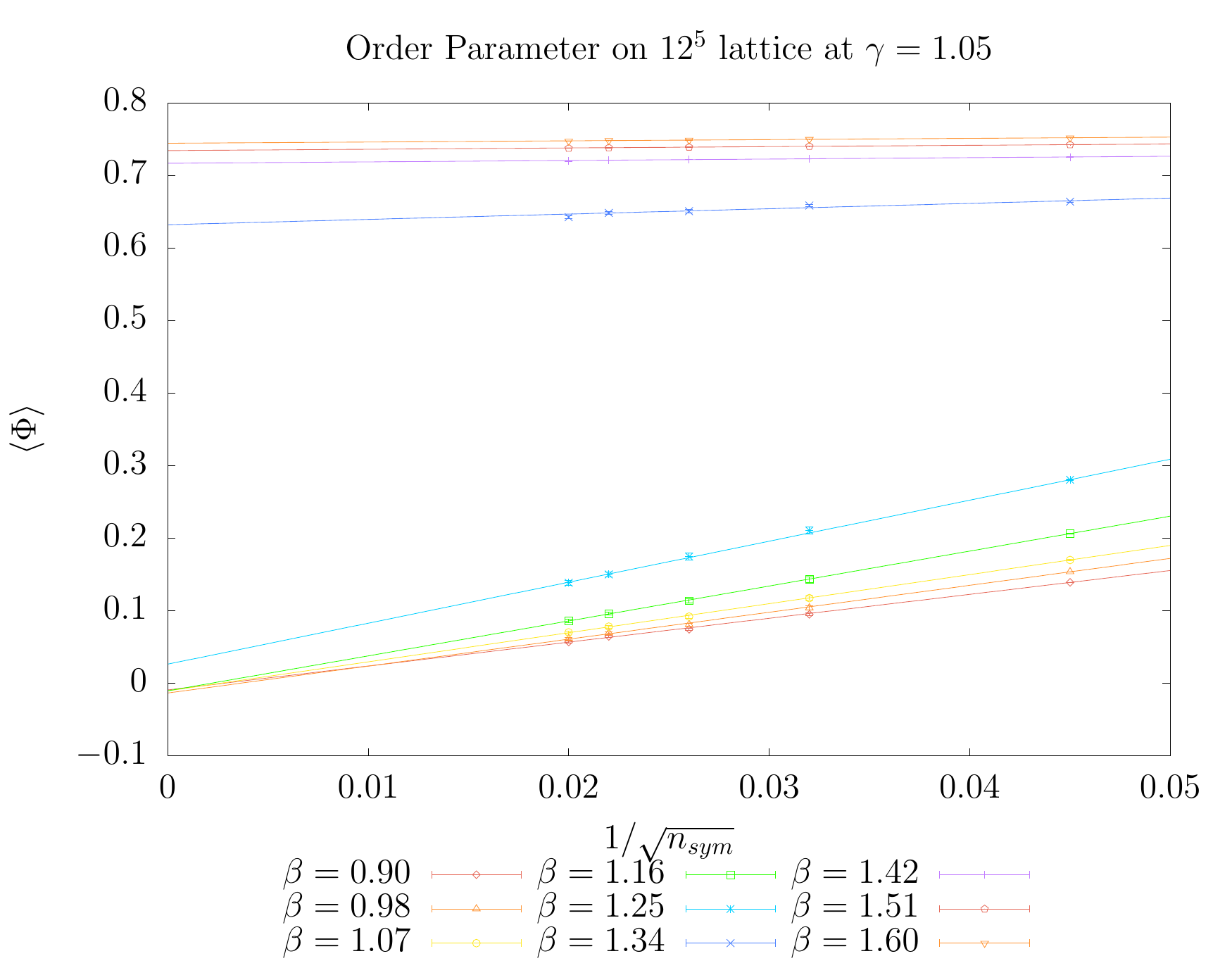}
  \label{PhicoinftoHiggs}
  }
 \subfigure[~]{
 \includegraphics[scale=0.5]{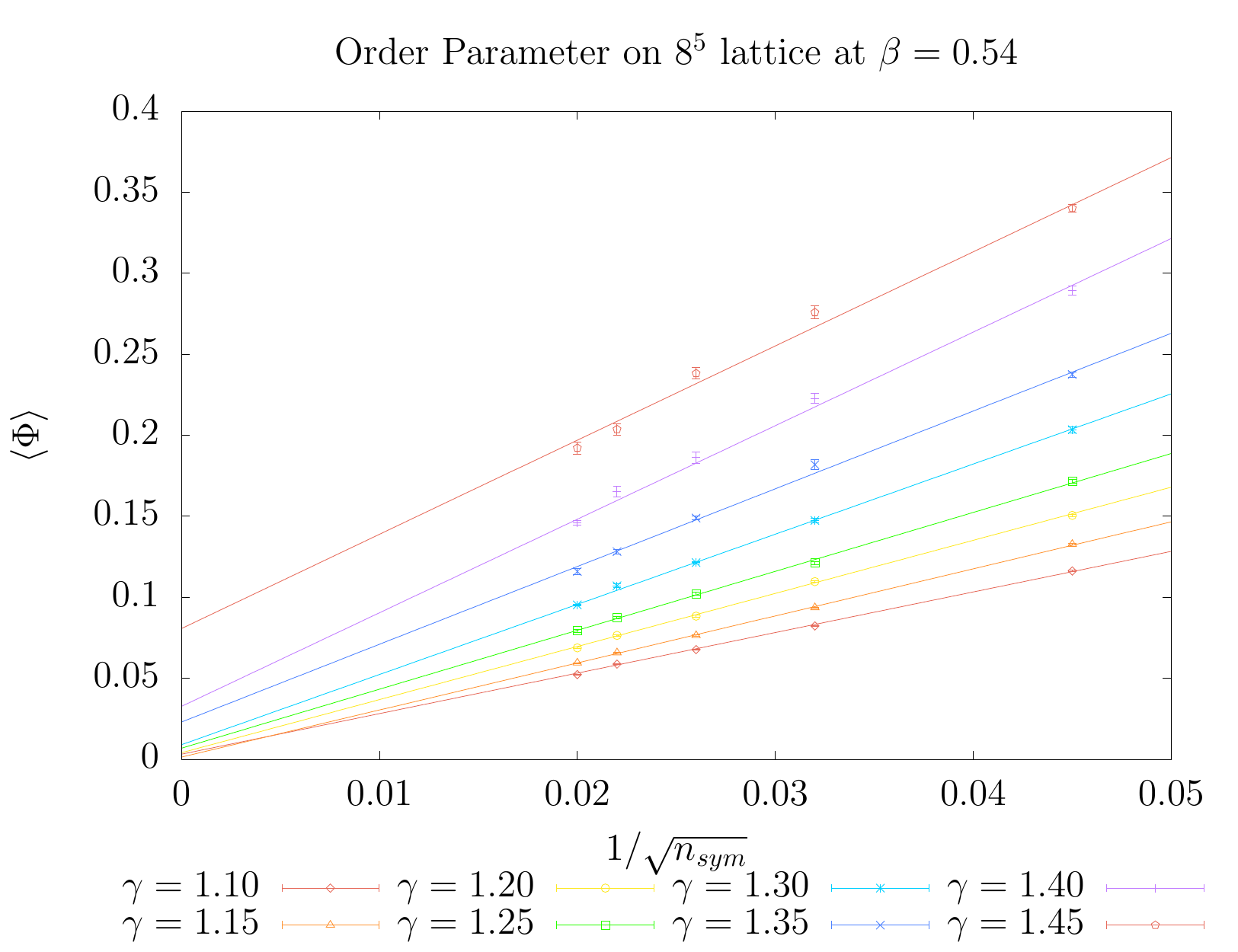}
  \label{nonthermo}
 }
 \caption{Symmetry-breaking transitions are located via extrapolation of data for the order parameter $\langle \Phi_{n_{sym}}\rangle$ vs.\ $1/\sqrt{n_{sym}}$ to $n_{sym} \ra \infty$.  Some sample data is shown.  (a) No transition from $\langle \Phi\rangle=0$ to a nonzero value is found as $\beta$ is varied, at fixed $\g=0.4$,
 across the confinement to massless phase transition.  Here the data at different $\b$ fall on the same line. (b) Symmetry breaking across the massless to Higgs transition.  Fixed $\b=2.1$, range of $\g$ values. (c) Symmetry breaking across the confinement to Higgs transition near the thermodynamic endpoint; fixed $\g=1.05$, range of $\b$ values.  (d) To the right of the endpoint of thermodynamic confinement-to-Higgs transitions there are still symmetry breaking transitions between the two phases, as illustrated in this data at fixed $\b=0.54$ at a range of $\g$ values.}
 \label{oparam}
 \end{figure*} 
 
     Now we explain the calculation of the Edwards-Anderson style order parameter, for the breaking of the global $Z_2$ center subgroup
of the gauge group.  Link and scalar field variables are updated in the usual way in update sweeps. However, the data-taking ``sweep'' of the lattice is not a single sweep, but actually a set of $n_{sym}$ update sweeps, in which the link variables on the $t=0$ time slice are held fixed,
while all other variables are updated as usual.  Then $\overline{\phi}(\vx)$ is the average value of $\phi(\vx,0)$ obtained at fixed
$U_i(\vx,t=0)$ over $n_{sym}$ sweeps, and the order parameter $\Phi_{n_{sym}}[U(0)]$ is computed from eq.\ \rf{Phi}.  An average
of all the data-taking values of $\Phi_{n_{sym}}[U(0)]$, each of which is taken at a different set of fixed link variables $U(\vx,0)$, gives
an estimate for $\langle \Phi_{n_{sym}} \rangle$.  On general statistical grounds we must have
\beq
        \langle \Phi_{n_{sym}} \rangle =  \langle \Phi \rangle +  {\text{const.} \over \sqrt{n_{sym}}} \ ,
\eeq
and we use this expression to extrapolate to the $n_{sym} \ra \infty$ limit, as seen in Figs.\ \ref{Phim0toconf} to \ref{nonthermo}.

    In this work $\langle \Phi_{n_{sym}}\rangle$ is computed for each choice of $n_{sym}$ in a separate run, with $n_{sym}$ ranging from 500 to 2500.  Each   $\langle \Phi_{n_{sym}}\rangle$ is evaluated from the average of 30 data taking procedures, separated by 100 ordinary update sweeps.  In some cases we look for the transition at fixed $\beta$ and varying $\gamma$, or vice versa keeping $\gamma$ fixed and varying $\beta$.

Fig.\ \ref{Phim0toconf} is a plot of $\langle \Phi_{n_{sym}} \rangle$ vs.\ $1/\sqrt{n_{sym}}$ at fixed $\gamma=0.4$ and a variety of $\b$, from
the confinement to the massless phase. The data points at different
$\b$ overlap one another, and in each case the order parameter extrapolates to zero.  The points at $\b<1.65$ are in the confined phase,
the points at higher $\b$ are in the massless phase, and we have already determined the thermodynamic transition separating the
massless from the confined phase.  So what we learn from this plot is that the Edwards-Anderson style order parameter 
$\langle \Phi \rangle=0$
in both the confined {\it and} the massless phases. Fig.\ \ref{Phim0toHiggs}  is a plot of $\langle \Phi_{n_{sym}} \rangle$ vs.\ $1/\sqrt{n_{sym}}$ at fixed $\b=2.1$ and a variety of $\g$.
This data shows the transition from $\langle \Phi \rangle=0$ in the massless phase to $\langle \Phi \rangle>0$ in the Higgs phase.
Checking the absence of a symmetry breaking transition from the massless to the confinement phase, and the presence of such a transition from the massless to the Higgs phase, was the primary motivation of our work, and the data appears to confirm the prediction.  Of course,
since $\langle \Phi \rangle=0$ in the confinement and massless phases, and $\langle \Phi \rangle > 0$ in the Higgs phase, it follows that there must be a symmetry breaking transition separating the confinement and Higgs phases.  Examples of the relevant data are shown in
Fig.\ \ref{PhicoinftoHiggs}. This figure displays the determination of the symmetry breaking transition point at $\g=1.05$, quite close to the thermodynamic endpoint, where we take data a number of 
$\b$ values.  At lower values of $\b$, in the confined phase, the data extrapolates, as expected, to $\langle \Phi \rangle=0$, and at higher values, in the Higgs phase, the data extrapolates to $\langle \Phi \rangle>0$, also as expected.  Fig.\ \ref{nonthermo} shows the determination of the symmetry breaking transition at $\b=0.54$, in a region where there are no nearby thermodynamic transitions.

   The transition point, whether in $\beta$ at fixed $\gamma$ or vice versa, is the coupling midway between two neighboring values, one
of which extrapolates (within errors) to zero at $n_{sym} \ra \infty$, and the other which extrapolates to a non-zero value.  For example,
suppose we vary $\gamma$, and among the set of trial gamma values, $\gamma_i$ is the highest coupling at which $\langle \Phi_{n_{sym}} \rangle$ extrapolates to zero, and $\gamma_{i+1}$ is the lowest value at which $\langle \Phi_{n_{sym}} \rangle$ extrapolates to a non-zero
value within errors.  Then we estimate ${\g=(\g_i + \g_{i+1})/2}$ as the transition point, with ${\d \gamma = (\g_{i+1} - \g_i)/2}$ as the error.
 
\

\begin{figure}[h]
\includegraphics[scale=0.5]{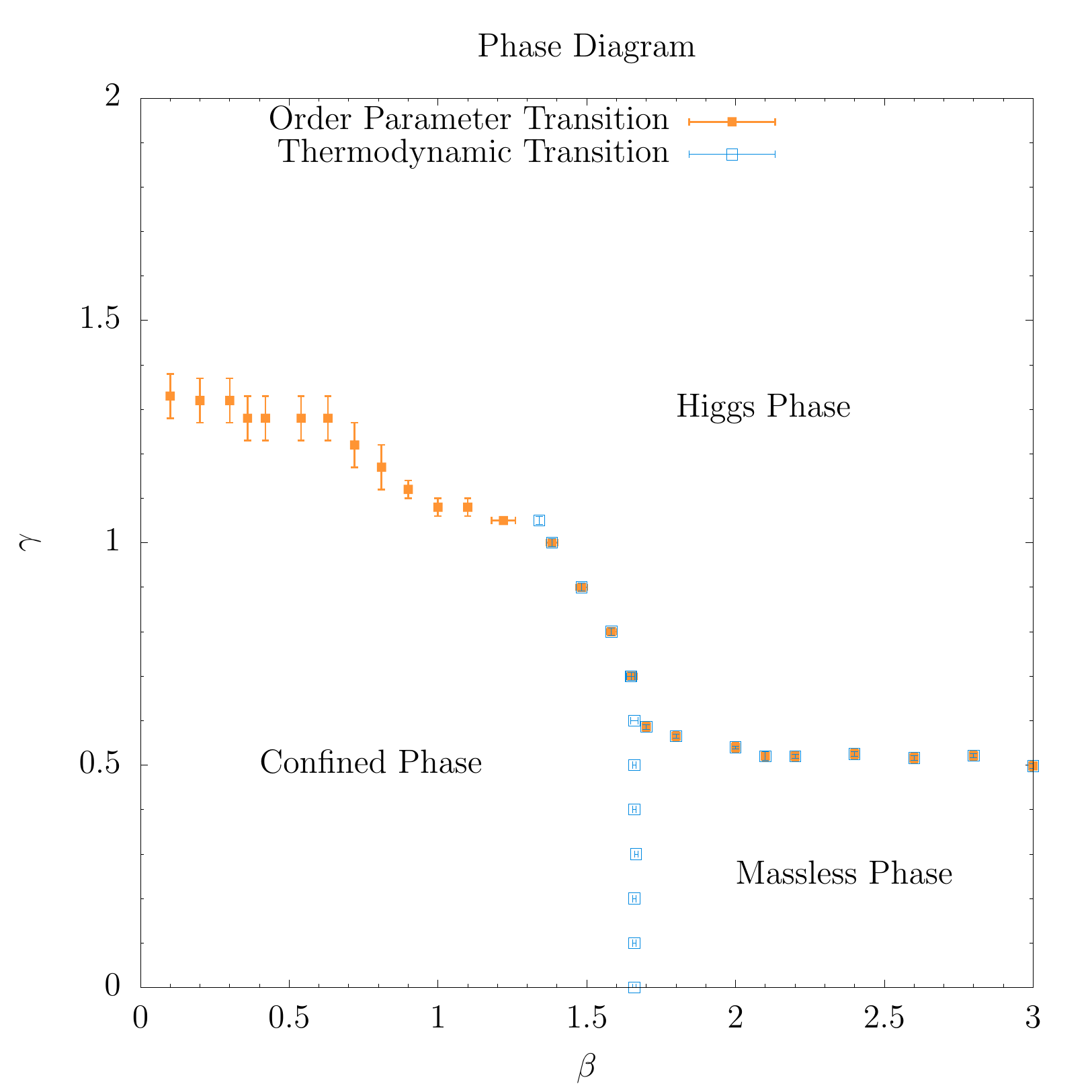}
\caption{Combined phase diagram of SU(2) gauge Higgs theory in $D=5$ dimensions, with thermodynamic transitions shown in blue, and symmetry breaking transitions, as detected by the $\langle \Phi \rangle$ order parameter, shown in orange.}
\label{phasecombo}
\end{figure}

    Proceeding in this way we can map out the regions of broken (Higgs) and unbroken (confinement and massless) custodial symmetry,
with the result shown in Fig.\ \ref{phasecombo}.  Here we show the combined data with the symmetry-breaking transition points shown in
orange, and points of thermodynamic transition, already seen in Fig.\ \ref{thermo}, shown in blue.  Note the coincidence of the thermodynamic and symmetry breaking transition points to the Higgs phase, at least to near the end of the thermodynamic transition line; the figure shows our best estimate for the position of that endpoint.  Because of critical slowing down in that region we have not obtained a reliable estimate for the position of the symmetry breaking transition near that endpoint.

    Above $\gamma \approx 1.35$  we find $\langle \Phi \rangle>0$ at all $\b$, with the Higgs phase extending all the way down
to $\b=0$.  So there is one line of transition from the confinement phase to the Higgs phase, roughly horizontal at $\b<0.6 $,  unaccompanied
by a thermodynamic transition.  Again this is expected from the Osterwalder-Seiler theorem \cite{Osterwalder:1977pc}; we already knew
that the confinement and Higgs phases could not be entirely isolated from one another by thermodynamic transitions.  It is however
encouraging that where there are thermodynamic transitions from the confinement to the Higgs phase these seem to coincide (with the possible exception of the thermodynamic endpoint) with the symmetry breaking transition points. \\

\section{Conclusions}

   An investigation of lattice SU(2) gauge Higgs theory in $D=5$ dimensions  confirms an important, and up to now untested, assertion of the
proposed identification of the Higgs phase as a phase in which the global center subgroup of the gauge group is spontanteously
broken.  This identification presupposes that the massless phase and the confinement phase are both phases of unbroken symmetry.
In fact, since the transition in the gauge-invariant order parameter $\langle \Phi \rangle$ is the only way to clearly localize the transition
between the confinement and Higgs phases, the best one can do is to study whether this transition line agrees with the points of
thermodynamic transition up to the end point of thermodynamic transition, and this agreement is seen.  Beyond that, there is no simple
independent check of the location of the transition (although a violation of the \Sc\ condition in the symmetric phase would falsify the identification of the symmetric phase with  \Sc\ confinement).    The massless phase, however, is entirely separated thermodynamically from the massive phases, so in this case there is a clear check of whether the order parameter vanishes in
the massless phase, remaining zero across the transition to the confining phase, and acquiring a non-zero expectation value in the
Higgs phase.  This behavior is clearly seen in our data, and provides some further support for the broken symmetry characterization
of the Higgs phase, as recently put forward in ref.\ \cite{Greensite:2020nhg}.

\acknowledgments{I would like to thank Jeff Greensite for his guidance in the course of this investigation. This work was supported by the US Department of Energy under Grant No.\ DE-SC0013682.}

\bibliography{sym3}
\end{document}